\documentclass[a4paper]{ar-1col}
\usepackage[numbers]{natbib}
\usepackage{amssymb}
\usepackage{hyperref}
\usepackage[english]{babel}

\jname{Xxxx. Xxx. Xxx. Xxx.}
\jvol{AA}
\jyear{YYYY}
\doi{10.1146/((please add article doi))}

\newcommand{\citex}[1]{\citep{#1}}
\newcommand{\pd}{\partial}
\newcommand{\ud}{\mathrm{d}}
\begin{document}

\markboth{B.~M\"uller}{Neutrinos from
Core-Collapse Supernovae}

\title{Neutrino Emission as Diagnostics of Core-Collapse Supernovae}

\author{B.~M\"uller$^1$
\affil{$^1$Monash Centre for Astrophysics, School of Physics and Astronomy, Monash University, Clayton, Australia, VIC~3800; email: bernhard.mueller@monash.edu}
}

\begin{abstract}
With myriads of detection events from a prospective Galactic core-collapse supernova, current and future neutrino detectors will be able to sample detailed, time-dependent neutrino fluxes and spectra. This offers enormous possibilities for inferring supernova physics from the various phases of the neutrino signal from the neutronization burst through the accretion
and early explosion phase to the cooling phase. The signal will constrain the time evolution of bulk parameters of the young proto-neutron star like its mass and radius as well as the structure of the progenitor, probe multi-dimensional phenomena in the supernova core, and constrain the dynamics of the early explosion phase. Aside from further astrophysical implications, supernova neutrinos may also shed further light on the properties of matter at supranuclear densities and on open problems in particle physics. 
\end{abstract}

\maketitle

\tableofcontents

\section{INTRODUCTION}
The  collapse and the ensuing explosion of massive stars 
as a core-collapse supernova constitute one of the most intriguing processes 
in astrophysics in which neutrinos play a crucial for  the 
dynamics of a macroscopic system, and one of the few  detectable
sources of neutrinos outside the solar system.

\subsection{Dynamics of Collapse and Explosion}
Many elements of this phenomenon are by now safely established 
by theory, and have even been corroborated to 
some degree by the ground-breaking detection of about two dozen 
neutrinos from SN~1987A in the Large Magellanic Cloud
\citep{hirata_87,bionta_87,alexeyev_88}.
The phase of collapse and bounce is now well understood
and has been discussed extensively in the classic paper
of  \citex{bruenn_85} and in other reviews \citex{bethe_90,janka_07}:
Once the iron core of
the progenitor star grows to roughly the Chandrasekhar
mass and has reached sufficiently high densities
by quasi-static contraction, electron captures
on heavy nuclei and photodisintegration of heavy nuclei
(the latter being more important for higher core entropy)
eventually lead to collapse on a dynamical time scale.
As the density and the electron chemical potential
increase, electron captures on heavy nuclei and
the few free protons that are present in NSE happen more rapidly and accelerate
\begin{marginnote}[]
\entry{NSE}{nuclear statistical equilibrium}
\end{marginnote}
the collapse. Initially, the electron neutrinos ($\nu_e$) produced
by the electron captures leave the core unimpeded, until
the neutrino mean free path becomes comparable to the core
radius at densities of a few $\mathord{\sim}10^{11}\, \mathrm{g}\, \mathrm{cm}^{-3}$ so that the emitted neutrinos are
trapped and further loss of lepton number from the core  (deleptonization)  ceases. This does not halt the collapse,
however, which only stops once the core density overshoots
nuclear saturation density and the repulsive nuclear
forces lead to a stiffening of the equation of state (EoS) and
an elastic rebound (``bounce'') of the homologous inner core.

In the wake of the bounce, a shock wave is launched at
the edge of the inner core. The shock quickly turns
into a stalled accretion shock as the initial energy
of the rebound is consumed by the disintegration
of heavy nuclei into free nucleons in the shock, and
by rapid neutrino losses once the shock reaches densities
of $\mathord{\sim}10^{11}\, \mathrm{g}\, \mathrm{cm}^{-3}$.
The position of the shock then adjusts quasi-statically
to the decreasing mass accretion rate, reaching a
maximum radius of $100\texttt{-} 200\, \mathrm{km}$ 
about $100\, \mathrm{ms}$ after bounce before slowly receding again.

How the stalled shock is then ``revived'' in most progenitors, 
i.e., made to propagate out dynamically to expel the outer 
layers of the star, remains the subject of active research 
\citep[see the reviews of][]{mezzacappa_05,janka_12,burrows_13,mueller_16b}.
The most promising scenario for the majority of core-collapse
supernovae is the delayed neutrino-driven mechanism \citep{bethe_85}. In the neutrino-driven paradigm
the shock is revived thanks to the reabsorption of
a fraction of the neutrinos emitted from the proto-neutron
star (PNS) surface in the gain region
\begin{marginnote}[]
\entry{PNS}{Proto-neutron star, i.e., the hot, and still relatively proton-rich compact remnant during the early seconds of a supernova that
later deleptonizes and cools to become a veritable
neutron star.}
\entry{gain region}{the region behind the shock where neutrino heating dominates over neutrino cooling.}
\end{marginnote}
behind the
shock. If neutrino heating is sufficiently strong, the
increase in thermal pressure pushes the shock outwards, which 
in turn increases the mass of the gain region and hence the 
efficiency of neutrino heating so that runaway shock expansion 
ensues. In all but the least massive progenitors
\citep{kitaura_06}, the neutrino heating needs to be supported 
by multi-dimensional (multi-D)  fluid instabilities like convective
overturn \citep{herant_94,burrows_95,janka_96} or the SASI instability \citep{blondin_03,foglizzo_07},
\begin{marginnote}[]
\entry{SASI}{standing accretion-shock instability}
\end{marginnote}
which 
manifests itself in the form of dipolar or quadrupolar
shock oscillations. One alternative to this scenario is the 
magnetorotational mechanism \citep[e.g.][]{akiyama_03,dessart_07_a,winteler_12,moesta_14}, which may explain the small
fraction of unusually energetic \emph{hypernovae} with explosion
energies of up to $\mathord{\sim}10^{52}\, \mathrm{erg}$,
but which requires rapidly rotating progenitors.
There are many indications, e.g.\ from the birth
spin periods of pulsars \citep{faucher_06} and asteroseismic 
measurements of core rotation in low-mass stars
\citep{mosser_12}, that such rapidly rotating
progenitors are rare, and that the core of massive
stars typically rotate slowly due to efficient angular
momentum transport in stellar interiors. Other mechanisms
have also been proposed, most notably the phase-transition
mechanism of \citep{sagert_09,fischer_18}, which involves a second
collapse and bounce of the PNS after a hypothetical 
first-order QCD phase transition, which launches another 
shock wave that is sufficiently powerful to explode the star.

\subsection{Neutrino Emission -- Rough Estimates and Scales}
Regardless of the supernova mechanism, neutrinos dominate
the energy budget of the supernova core 
and carry away most of the energy liberated by gravitational
collapse, which is essentially equal to the binding energy $E_\mathrm{bind}$ of the young neutron star. 
$E_\mathrm{bind}$ 
is of the order of $GM^2/R$ in terms of the (gravitational) neutron star mass $M$ and radius $R$; a more precise fit to
solutions of the stellar structure equations yields
\citep{lattimer_01}
\begin{equation}
\label{eq:ebind}
  E_\mathrm{bind}\approx
  0.6 \frac{G M^2}{R}
  \left(1-\frac{1}{2}\frac{G M}{Rc^2}\right)^{-1}
.
\end{equation}
Because of neutrino trapping, this energy is radiated
away only on time scales of seconds with total
luminosities of all flavours of $\mathord{\sim}10^{53}\,\mathrm{erg}\, \mathrm{s}^{-1}$.
As the neutrinos decouple from the matter only at the 
``neutrinosphere'' at the PNS surface during  
the first few seconds of its life,  their emerging spectrum 
reflects an environment with a  temperature of a few 
$\mathrm{MeV}$ rather than tens of $\mathrm{MeV}$ in the 
PNS interior. Together with the
radius of the neutrinosphere, the PNS
surface temperature sets the scale for the luminosity
according to the Stefan-Boltzmann law
\begin{equation}
L_\mathrm{\nu}\sim 4\pi \sigma_\mathrm{fermi} R^2 T^4,
\end{equation}
where $\sigma_\mathrm{fermi}=                                                         
4.50 \times 10^{35} \, \mathrm{erg}\, \mathrm{MeV}^{-4} \mathrm{s}^{-1} \mathrm{cm} ^{-2}$ is the radiation constant for massless fermions with
vanishing degeneracy.

Based on such simple considerations, the detection of
the  neutrinos from SN~1987A \citep{hirata_87,bionta_87,alexeyev_88} 
was already sufficient to validate the basic theoretical
picture of core collapse. The total count, energy,
and timing of the detected neutrinos established
that a compact object with a binding energy of
a few $10^{53}\, \mathrm{erg}$ (assuming equipartition between flavors) 
was formed and emitted neutrinos for a few seconds from a surface
region with a radius of tens of $\mathrm{km}$ and a temperature of a
few $\mathrm{MeV}$ \citep{bruenn_87,bahcall_87,burrows_88a,arnett_89}.

The neutrino signal from a prospective Galactic supernova could provide
 considerably more information on the dynamics in the supernova core, 
the progenitor, and on problems in nuclear and particle physics. 
With current and future instruments, the principal difference to the
case of SN~1987A would consist in better statistics,
which would provide detailed time-dependent
fluxes for the arriving $\bar{\nu}_e$ and to a lesser extent
the $\nu_e$, allow for
a much better determination of the neutrino energy spectrum,
and constrain the flux of heavy-flavor neutrinos to some degree.

Some excellent reviews on the supernova neutrino signal have been written in recent years and may
also be consulted for further reference. This particular review seeks to fill
the middle ground with less of a focus on the basic physical principles
(neutrino transport, weak interaction rates, etc.) and a
broader coverage of the diagnostic potential of the neutrino signal than 
\citep{janka_handbook}, but a more selective and compressed 
approach than the very extensive reviews of \citex{mirizzi_16,horiuchi_18}.

\section{PREPARING FOR THE NEXT GALACTIC SUPERNOVA}

\subsection{Prospects for Supernova Neutrino Detection}
Current and future supernova neutrino detectors employ different
detector materials and detection principles, and will complement
each other in the event of a Galactic supernova.

Water Cherenkov detectors can accommodate large
detector volumes, and will have the highest count rates.
They are primarily sensitive to $\bar{\nu}_e$ via the 
\begin{marginnote}[]
\entry{IBD}{inverse $\beta$-decay}
\end{marginnote}
IBD reaction
$\bar{\nu}_e+p\rightarrow n+e^+$. ``Classical'' water
Cherenkov detectors are capable of measuring the energies
of detected MeV neutrinos; examples include the operational  SuperKamiokande (SuperK)
detector \citep{ikeda_07} with $\mathord{\sim}10,000$ events for a Galactic supernova at a typical
distance of $10\,\mathrm{kpc}$ and its planned successor
Hyper-Kamiokande (HyperK; \citealp{hyperk_13}) with $\mathord{\sim} 250,000$ events. Particularly
large detector volumes can be realized in long-string water
Cherenkov detectors like IceCube \citep{abbasi_11}. However, MeV neutrinos
will only be detected through an increase in the dark current in such
detectors, and no energy information will be available. The primary advantage of IceCube for supernova
neutrino detection is its excellent
time resolution and high total event count of $10^5\texttt{-}10^6$ events.

In liquid scintillator detectors, IBD is also the primary
detection channel, but since they are are limited to smaller volumes,
the expected count rates are smaller than for HyperK with
$\mathord{\sim}15,000$ and $\mathord{\sim}5,000$ IBD events for the future
JUNO \citep{an_16} and LENA \citep{wurm_12}  detectors, respectively.
However, they offer excellent energy resolution and allow
for the reconstruction of the $\nu_e$ signal to some degree.
Liquid scintillator detectors currently in operation
(KamLAND, Borexino, Baksan, etc.) will only detect a few
hundred events unless the supernova is exceptionally close.
The NOvA detectors also have a sufficiently large volume to observe
a few thousands events, but are geared
towards GeV neutrinos; work on supernova
neutrino detection with these instruments is in progress \citep{vasel_17}.

Liquid argon detectors provide the best
handle on the $\nu_e$ signal through the reaction
$\nu_e+{}^{40}\mathrm{Ar}\rightarrow
{}^{40}\mathrm{K}+e^-$. With a detector
mass of $40\, \mathrm{kt}$, the future DUNE detector 
\citep{acciarri_16}
will  provide good sampling of
the $\nu_e$ light curves with
$\mathord{\sim}3,000$ events for a supernova at $10\,\mathrm{kpc}$.

Heavy-flavor neutrinos
(henceforth denoted as $\nu_x$) will only be detected via neutral-current
scattering events, primarily neutrino-electron scattering
in water Cherenkov detectors, and also neutrino-proton scattering
in liquid scintillator detectors. Although future detectors
will measure a sizable number of scattering events
(e.g.\ a few thousand in LENA), the reconstruction of the heavy-flavor neutrino flux
is not trivial: The  $\nu_x$
are conflated with 
$\nu_e$ and $\bar{\nu}_e$ in the scattering channel, and
the exact energy of the scattered neutrino cannot be reconstructed.

For more information and other detectors types
we refer to dedicated reviews
on supernova neutrino detection \citep{scholberg_12,mirizzi_16}

\subsection{Neutrino Signal Predictions: Theoretical Challenges and Uncertainties}

There is a flipside to the prospect of accurate,
 time-dependent measurements of supernova neutrino fluxes and spectra: 
Different from the historic example of SN~1987A, 
uncertainties 
in the predicted neutrino emission on the level of a few
percent or more can become relevant for inferring physical parameters. Such uncertainties concern
various aspects of the supernova problem, e.g., 
numerical
approximations for neutrino transport and neutrino-matter interaction rates. These cannot be treated
at length here, and we instead refer the reader to the literature. Strengths and weaknesses of currently
employed methods for neutrino transport are
discussed in the reviews of \citex{mueller_16b,janka_16},
and in recent years a number of papers have helped to gauge uncertainties in the modelling by code comparisons
\citep[e.g.][]{messer_98,liebendoerfer_05,mueller_10,just_18,kotake_18,oconnor_18,pan_19} and by investigating
variations in the neutrino interaction rates
\citep[e.g.][]{buras_03,buras_06a,lentz_12a,lentz_12b,martinez_12,mueller_12a,bartl_16,bollig_17,horowitz_17,fischer_18}.

There are also unresolved problems concerning 
neutrino flavor conversion in supernova cores that translate into
uncertainties in the observable fluxes in the different flavors.
Specifically, research on collective oscillations is still very much
in a state of flux so that we can only outline the problem
and refer to \citex{duan_10,mirizzi_16}
for more detailed overviews.

 Flavor conversion in supernovae is determined by the interplay
of three different types of terms in the neutrino Hamiltonian.
The vacuum terms and the matter terms that arise from neutrino
forward scattering on charged leptons give rise
to the two familiar Mikheyev-Smirnov-Wolfenstein (MSW) resonances
\citep{wolfenstein_78,mikheyev_85} at densities
of $\mathord{\sim}10^3 \, \mathrm{g}\, \mathrm{cm}^{-3}$
(H-resonance) and
$\mathord{\sim}10\, \mathrm{g}\, \mathrm{cm}^{-3}$
(L-resonance). The effect of the MSW resonances
alone is rather well understood; with the three neutrino flavors
in the standard model, the outcome depends on the (unknown) mass hierarchy
and the structure of the star in the resonance regions (adiabatic vs.\ non-adiabatic conversion).
As a rule of thumb, MSW flavor conversion in the
normal mass hierarchy alone would result
in a complete swap of
$\nu_e$ with heavy-flavor neutrinos and a large
survival rate of 0.68 for  $\bar{\nu}_e$ for most progenitors
during the early signal phase; for an inverse mass hierarchy, 
the survival rates of $\nu_e$ and $\bar{\nu}_e$ would be 
0.32 and 0, respectively.
Some refinements of this picture will be discussed in Sections~\ref{sec:burst} and \ref{sec:exotic}.

However, flavor conversion in supernovae is complicated by
the high neutrino number densities in the environment of
the PNS. Under these conditions, the terms
for neutrino-neutrino forward scattering (``neutrino self-refraction'')
in the Hamiltonian can no longer be ignored \citep[e.g.][]{pantaleone_92,pastor_02} and drive
collective flavor conversion of neutrino and antineutrinos
that conserves lepton family number. The self-refraction
terms turn  flavor conversion into a non-linear problem
with an extremely complex phenomenology that is not yet fully understood.
Additional flavor conversion modes have appeared whenever
new dimensions -- such as the angular distribution of neutrinos in phase space --
were added to the problem \citep[e.g.][]{sawyer_09,raffelt_13b,mirizzi_13,mirizzi_15,dasgupta_15}, and the numerical treatment is ripe with
pitfalls that give rise to spurious instabilities \citep{sarikas_12c,morinaga_18}.
During the pre-explosion phase, the matter terms may be sufficiently large
to suppress collective flavor conversion \citep{chakraborty_11a,sarikas_12a}, but the final verdict
on the conditions and outcome of these collective oscillations is still pending.

\section{COLLAPSE AND NEUTRINO BURST}
\label{sec:burst}
Even before the onset of collapse, a supernova progenitor
is already a strong source of MeV-neutrinos that mostly
come from thermal emission processes during advanced
burning stages. For nearby supernovae, the neutrino emission
from the core silicon burning stage may be detectable as the first
signature of the impending collapse
by future liquid scintillator detectors
like JUNO and DUNE, or already by SuperK if doped with gadolinium
\citep{odrzywolek_04,kato_15,yoshida_16,asakura_16,patton_17}.
This would not only provide an advance warning
for the supernova, but could also serve as a diagnostic
for the progenitor mass and even reveal the timing
of some of the final core and shell burning episodes
\citep{yoshida_16}. Improvements are still
needed, however, to gauge the full diagnostic
potential of pre-supernova neutrinos, for
example by a more rigorous treatment of
the emission processes, including $\beta$-processes \citep{patton_17}.

\begin{figure}
    \centering
    \includegraphics[width=\linewidth]{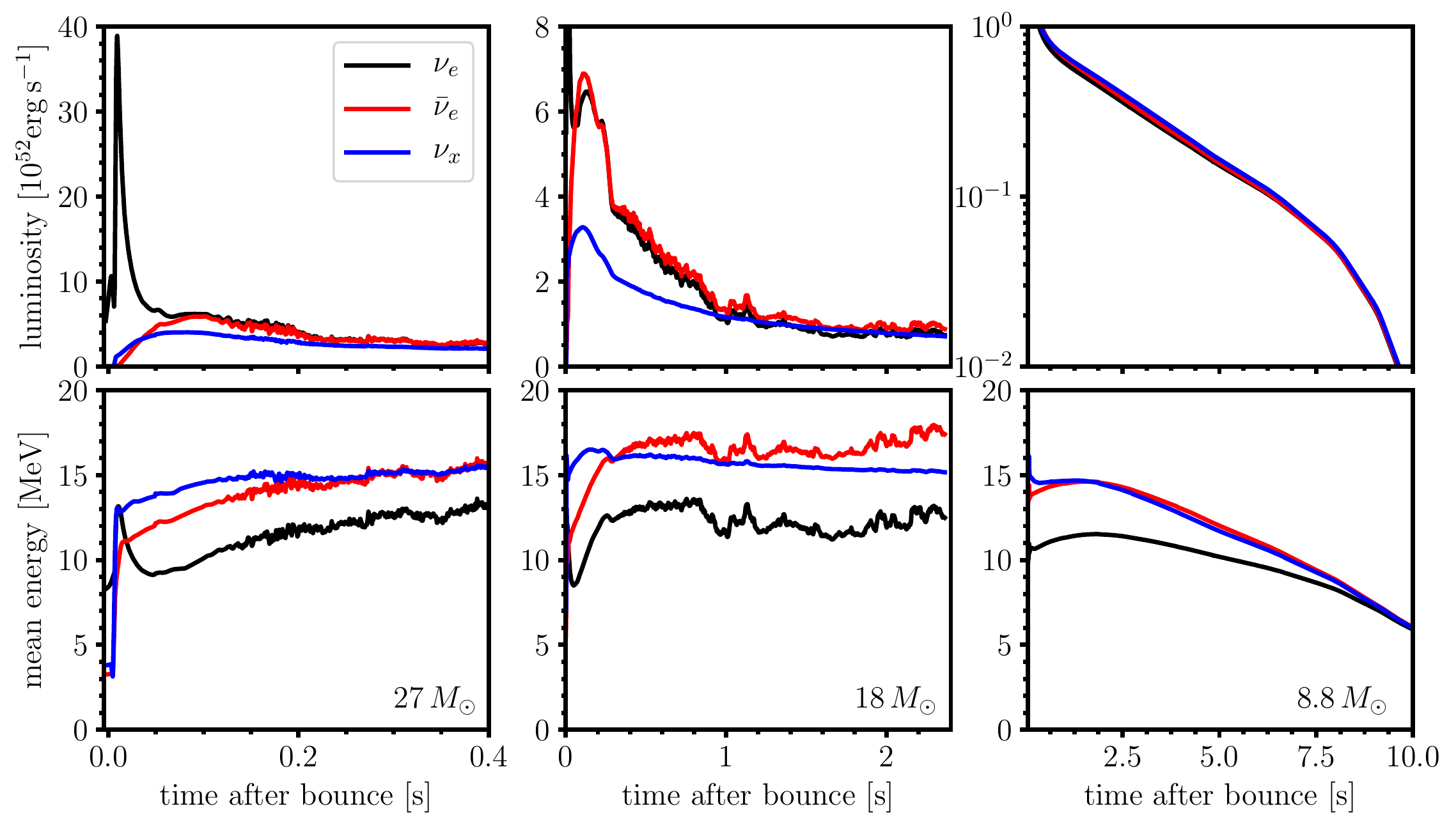}
    \caption{Neutrino luminosities
    and mean energies from different simulations. The
    2D model of an $27 M_\odot$
    star (\citealp{mueller_14}; left column) illustrates the burst phase, the
    accretion phase, and the early explosion phase with the characteristic excess in
    the luminosity of $\nu_e$ and $\bar{\nu}_e$. The 
    3D explosion model of an $18 M_\odot$
    star (\citealp{mueller_17}; middle) extends
    further into the explosion phase and
    shows the luminosities of different flavors moving closer to equipartition as accretion subsides. The $8.8 M_\odot$
    electron-capture supernova
    model from \citep{mueller_10} (right column) 
    shows the Kelvin-Helmholtz cooling phase with good equipartition and a visible
    decline of neutrino mean energies after
    $\mathord{\sim}1.5 \, \mathrm{s}$.}
    \label{fig:neutrino_signal}
\end{figure}

When the collapse of the iron core starts in earnest, the production
of $\nu_e$ by electron captures on heavy nuclei and the few
free protons available in NSE becomes
the dominant source of neutrinos. The $\nu_e$-luminosity increases  to about $10^{53}\, \mathrm{erg}\, \mathrm{s}^{-1}$ around the time of trapping, and the mean energy climbs to  $\mathord{\sim}10\,\mathrm{MeV}$. Trapping then leads to a small
dip in the luminosity as neutrino leakage is 
confined to a narrow region around the newly formed
neutrinosphere.
Neutrino emission again increases rapidly
after core bounce as the newly formed shock propagates
into regions of sufficiently low density and reaches
the neutrinosphere. Due to shock heating and low optical
depth, neutrinos are swiftly released in copious amounts
from the shocked matter, mostly via electron
captures on free protons. The emission of $\nu_e$
dominates by far since the electron fraction
$Y_e$ in the shocked matter is still relatively high
and far above the $\beta$-equilibrium value,
with $\nu_e$-luminosities transiently reaching
$3.5 \times 10^{53}\, \mathrm{erg}\, \mathrm{s}^{-1}$ in 
what is known as the neutronization burst
(see \textbf{Figure~\ref{fig:neutrino_signal}}
for typical light curves and neutrino mean
energies during the first seconds of a supernova).

The burst has substantial diagnostic value since its shape
is quite robust with little dependence on the progenitor
mass or the nuclear EoS \citep{kachelriess_05}:
With a megaton water Cherenkov detector like HyperK,
the burst could be used as a ``standard candle'' for a distance
determination within $\mathord{\sim}5\%$.
Moreover, the observation or non-observation
of the  $\nu_e$-burst constrains
the mass hierarchy, which determines
the $\nu_e$ survival probability in
the MSW resonance regions. In the normal mass hierarchy
the burst neutrinos would leave the star in the third mass
eigenstate and thus hit detectors on Earth with only a tiny overlap
with the $\nu_e$ flavor eigenstate.

As pointed out by \citex{serpico_12}, another independent handle on the mass hierarchy,  could come from
the signal of $\bar{\nu}_e$ and heavy flavor neutrinos.
Although the  $\nu_e$-emission dominates during the burst,
the light curves of 
$\nu_x$ begin to rise during the burst as thermal
emission processes (electron-positron pair annihilation,
bremsstrahlung, and neutrino pair conversion)
become important in the shock-heated matter. The emission of $\bar{\nu}_e$ 
by charged-current processes is inhibited as long as the matter is still
more proton-rich than in $\beta$-equilibrium, so that
the $\bar{\nu}_e$ light curve rises more slowly than that of
$\nu_x$. Since the mass hierarchy determines how the
emission of $\bar{\nu}_e$ and $\bar{\nu}_x$ in the supernova
core translates into $\bar{\nu}_e$ and $\bar{\nu}_x$
after the neutrinos undergo MSW flavor conversion at
radii of tens of thousands of kilometres, the 
fast or slow rise of the detected
$\bar{\nu}_e$ signal on Earth would point to an inverted
or normal mass hierarchy, respectively \citep{serpico_12}.

For a special core-collapse supernova channel arising
from super-asymptotic giant branch (SAGB) stars, flavor conversion during this early phase of neutrino emission works in a distinctly
different manner, which implies that a neutrino detection
could provide a smoking gun for such SAGB progenitors. These progenitors
are low-mass stars with a ZAMS mass
\begin{marginnote}[]
\entry{ZAMS}{Zero-age main sequence mass}
\end{marginnote}
around $8M_\odot$ (for single stars), which do not
go through all the hydrostatic burning stages up to the
formation of an Fe core, but undergo  collapse due to  electron captures on $^{20}\mathrm{Ne}$ and $^{24}\mathrm{Mg}$ in a highly degenerate O-Ne-Mg core \citep{nomoto_84,nomoto_87,jones_13}, and then explode
as  ``electron-capture supernovae'' with low explosion
energies \citep{kitaura_06}. They
exhibit a very steep density gradient outside the degenerate core, which
 moves the MSW resonances relatively close together, makes
the MSW flavor conversion non-adiabatic, and gives a larger role
to non-linear collective neutrino interactions because of the
low electron number densities. 
As shown in \citex{duan_08_b}, the emerging neutrino spectra exhibit
a spectral swap at $11\texttt{-} 15 \, \mathrm{MeV}$
(depending on the mass hierarchy) with a survival
probability of $\mathord{\sim} 0.68$
for $\nu_e$ of higher energies as these neutrinos
jump to the first mass eigenstate. If such a high survival
probability is measured for the burst neutrinos (which
presupposes that the distance to the supernova can be inferred
by other means), this would furnish direct proof for an SAGB progenitor.
The later phases of the signal could bolster such a progenitor
determination further as shown by \citex{lunardini_08} for
the case of the normal mass hierarchy: As the shock hits
the MSW resonances and the density gradients become shallower,
the MSW conversion becomes more adiabatic so that the
$\nu_e$ survival probability essentially drops to zero about
$100 \, \mathrm{ms}$ after bounce.

\begin{textbox}[htb]
\section{Electron-capture Supernovae}
The collapse of a star with an O-Ne-Mg core due to electron captures
is a channel towards core collapse that is still poorly
understood. Whether a star that has
undergone carbon core burning and evolves into an SAGB star can eventually
collapse and explode as an electron-capture supernova (ECSN) hinges
on many uncertainties regarding mass loss, mixing processes, and
turbulent flame propagation and nuclear physics after off-center
O ignition \citep{jones_13,jones_16,doherty_17,nomoto_17}.
This progenitor channel is likely very narrow for single stars
of solar metallicity \citep{poelarends_08,doherty_17}, but may be
wider at lower metallicity and in interacting binaries \citep{poelarends_08}.
To date, no observed transient has been unambiguously identified
as an ECSN, although various candidates
have been proposed, including the historic Crab supernova \citep{nomoto_82,smith_13},
SN~2008S \citep{botticella_09}, SN~2005cs \citep{pastorello_09}, and the subclass of type~IIn-P supernova \citep{smith_13}
with narrow emission lines.
Even for a Galactic ECSN,
uncertainties in the envelope structure and the presence of circumstellar material
may complicate the interpretation of the electromagnetic transient, and
a smoking gun for an SAGB progenitor from the neutrino signal would be most
valuable.
\end{textbox}

The $\nu_e$-burst and the rise phase
of the $\bar{\nu}_e$- and $\nu_x$-signal
also allow for a precise timing of the bounce.
The survival of the $\nu_e$-burst after oscillations
is not critical for this; assuming normal mass ordering (so that
the $\nu_e$-burst would not be seen in
liquid Argon detectors), IceCube will still be able to pinpoint the bounce to about $3.5 \, \mathrm{ms}$
using the rise of the measured $\bar{\nu}_e$-flux
\citep{halzen_09}.
This is also of relevance
in the context of concurrent neutrino and gravitational wave
detections. The neutrino signal is of  utility for
gravitational wave detection as it helps define the period
of interest for a signal search in a noisy data stream
\citep{yozokawa_15}. If there are correlated features in the neutrino
and gravitational wave signal, this can be exploited
to improve parameter estimation, and the bounce of
rotating progenitors provides the prime example for this. 
In this case there will be a strong gravitational
wave signal from the bounce of the rotationally deformed
core \citep[e.g.][]{dimmelmeier_08}, which is roughly
coincident with the neutrino burst. This temporal
correlation can be used to more accurately
determine the time of bounce and the degree
of rotation \citep{yozokawa_15}. For sufficiently rapid
rotation, the early neutrino signal also shows temporal
modulations, whose frequency is set by the fundamental
quadrupole mode that dominates the gravitational wave
spectrum \citep{ott_12}. However, even with HyperK
and for the most rapidly spinning PNSs,
these modulations in the neutrino signal will only
be detectable to $\mathord{\sim}1 \, \mathrm{kpc}$ 
according to the analysis of \citex{ott_12}.

\section{THE SIGNAL FROM THE ACCRETION AND EARLY EXPLOSION PHASE}
\label{sec:accretion}
\begin{figure}
     \centering
     \includegraphics[width=0.8\linewidth]{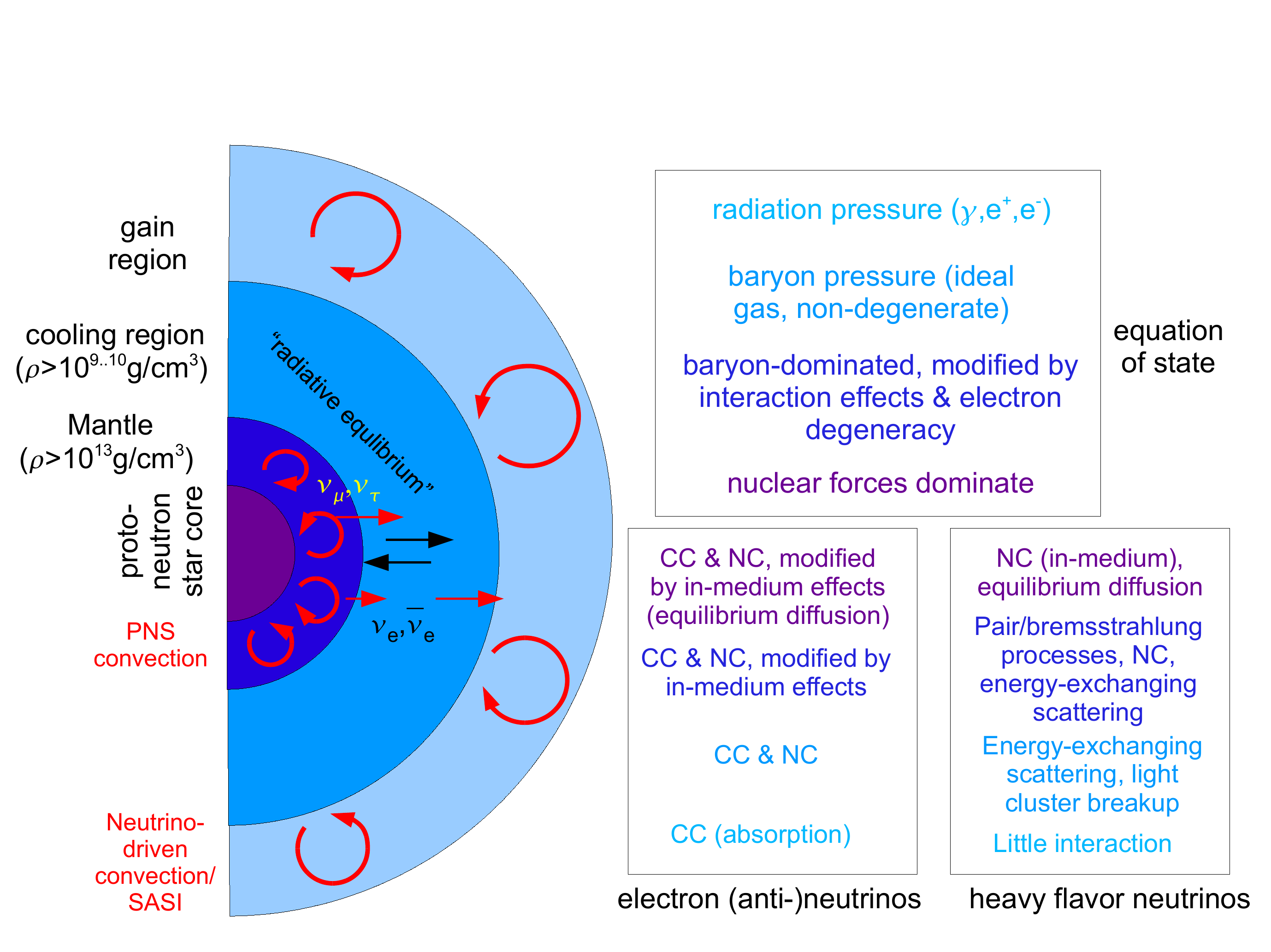}
     \caption{Sketch of the various regions in the supernova
     core, the EoS and transport regimes, and the neutrino interaction
     processes that are most relevant for the dynamics and
     the observable neutrino signal (neglecting flavor oscillations).
     (CC: charged-current, NC: neutral-current).}
     \label{fig:sketch}
 \end{figure}

Over time scales of tens of milliseconds, the supernova core develops a characteristic structure during the pre-explosion phase that is illustrated in \textbf{Figure~\ref{fig:sketch}}: 
The accretion
shock sits at a radius of $100 \texttt{-} 200 \, \mathrm{km}$,
and below it there is an extended ``hot-bubble region''
of high entropy. The EoS in this region
is dominated by photons and electron-positron pairs, 
and heating by
neutrinos from further inside dominates over neutrino cooling.
At moderately high densities 
of $\mathord{\sim}10^{10} \, \mathrm{g} \,\mathrm{cm}^{-3}$ to a few $10^{13} \, \mathrm{g} \,\mathrm{cm}^{-3}$ further inside, the pressure
is mostly provided by baryons and is roughly
described by an ideal gas law
$P \propto \rho T$. Close to the transition
between these two EoS regimes, neutrino cooling starts to
dominate over heating. Further inside in the
core of the PNS, the EoS is dominated
by nuclear interactions. Because of  shock heating
in the layers outside the small inner core 
of $\mathord{\sim} 0.5 M_\odot$
\citep{langanke_03} that remained in homologous collapse until bounce, the maximum temperature is reached off-center in an extended
mantle of moderate entropy between the core and the surface.

\subsection{Constraining Parameters of the Proto-Neutron Star and the Accretion Flow}
\label{sec:bulk_parameters}
The emission of electron flavor and heavy flavor
neutrinos is distinctly different in this environment.  
For all flavors, there is a diffusive flux to the PNS surface region driven by gradients in temperature and neutrino chemical potential. This diffusive flux is essentially determined
by the temperature and radius 
of the decoupling region (neutrinosphere) near the
PNS surface, i.e., by bulk parameters
of the PNS. This component of the neutrino
flux can be well described by the gray-body emission law
\begin{equation}
\label{eq:lum_x}
L_\mathrm{diff}= 4\pi \phi \sigma_\mathrm{fermi} R^2 T^4,
\end{equation}
where the grayness factor $\phi$ accounts for
the deviation from the Stefan-Boltzmann law.
Equation~(\ref{eq:lum_x}) adequately describes
the heavy-flavor luminosity, with the greyness parameter $\phi$
varying in the range $0.4 \texttt{-} 0.6 $ during the pre-explosion
and early explosion phase \citep{mueller_14}.

The emission of electron flavor neutrinos is not
only fed by the diffusion of thermal energy from the PNS core, but also by accretion energy. As the accreted matter settles
onto the PNS surface,
it enters into radiative equilibrium with deeper layers 
and must undergo net neutrino cooling to maintain a roughly constant temperature 
as it is compressed to higher densities. 
As heavy-flavor neutrinos  can only be produced at high densities 
$\gtrsim 10^{13} \, \mathrm{g} \, \mathrm{cm}^{-3}$
by pair creation, nucleon bremsstrahlung \citep{hannestad_98},
and neutrino pair conversion \citep{buras_03} and not by charged-current
processes in the more dilute atmosphere
\footnote{Charged-current production
of $\nu_\mu$
becomes relevant in the interior
of the PNS
due to the high temperatures and
chemical potentials of
electrons and $\nu_e$
\citep{bollig_17}.}, cooling proceeds mostly
by emission of $\nu_e$ and $\bar{\nu}_e$.
Only about half of the accretion energy
actually goes into $\nu_e$ and $\bar{\nu}_e$ 
\citep{mueller_14,huedepohl_phd,mirizzi_16},
since the accreted matter does not cool below a radiative equilibrium temperature
of several $\mathrm{MeV}$. 

Accounting both
for the diffusive component and the accretion luminosity, the luminosities
$L_{\nu_e}$ and $L_{\bar{\nu}_e}$ of electron neutrinos
and antineutrinos are well described by
\citex{huedepohl_phd,mirizzi_16}
\begin{equation}
\label{eq:lum_e}
L_{\nu_e}+L_{\bar{\nu}_e}=
2\beta_1 L_{\nu_x}+
\beta_2 \frac{G M \dot{M}}{R},
\end{equation}
where $M$ is the PNS mass, $\dot{M}$ is the mass
accretion rate, and $\beta_1=1.25$ and $\beta_2=0.5$ 
\citep{huedepohl_phd,mirizzi_16} are non-dimensional parameters.
The fact that $\beta_1\neq 1$ reflects that the gray-body contribution
need not be the same for electron flavor neutrinos and heavy-flavor neutrinos
because the decoupling of heavy-flavor neutrinos from the matter works 
differently due to the absence of charged-current reactions at low-densities.

Both $\nu_e$ and $\bar{\nu}_e$ carry roughly half of the electron flavor
luminosity. The exact split between $\nu_e$ and $\bar{\nu}_e$
is sensitive to the detailed neutrino interaction rates
 --- for example to the effect of nucleon potentials
on the charged-current rates \citep{martinez_12,roberts_12c},
especially when the accretion rate drops
during the explosion phase --- and to the structure of the neutrinospheric region and
the PNS convection zone below it \citep{buras_06b}.

During the accretion phase and early explosion phase, 
the high-density EoS primarily influences the neutrino emission indirectly
via the PNS radius and surface temperature in
Equations~\ref{eq:lum_x} and \ref{eq:lum_e}. Equations of state that yield
more compact PNSs result in higher neutrino luminosities
and mean energies \citep{janka_07,oconnor_13}. Over short time scales of hundreds of milliseconds,
diffusion is too slow to transport significant amounts
of energy and lepton number from the high-density core to the neutrinosphere,
and hence the precise transport coefficients and thermodynamic
properties well above saturation density have little direct effect on the neutrino signal.
Even the heavy flavor emission comes mostly from the extended
mantle rather than from the core during the pre-explosion phase.
This is not to say that nuclear interactions of the matter are unimportant
for the neutrino emission during this phase, since
they already affect the thermodynamic properties, composition, and
transport coefficients well below nuclear saturation density, e.g.\ through
correlation effects on the neutrino opacities \citep{burrows_98,burrows_99,horowitz_17}.
Many of the more recent  corrections in the 
treatment of such in-medium effects  typically affect the neutrino luminosities and mean energies on the level of a few percent \citep{horowitz_17,bollig_17}, which may be of smaller relevance in the context of neutrino observations, but can play an important role for shock revival
\citep{bollig_17}.

Like the luminosities, the neutrino spectra carry information on the
thermodynamic properties of the decoupling region.
For $\nu_e$ and $\bar{\nu}_e$, the emerging spectra
roughly reflect the energy-dependent equilibrium intensities
at the neutrinosphere \citep{janka_89a,janka_89b}. Because of the neutron-rich
conditions at the PNS surface, the opacity for
the absorption of $\bar{\nu}_e$ by protons is smaller than for 
the absorption of $\nu_e$ on neutrons, so that $\bar{\nu}_e$
decouple at smaller radii and higher temperatures.
The mean energy of $\bar{\nu}_e$ is thus higher by
about $2.5\, \mathrm{MeV}$. The precise difference
in mean energy is sensitive to the microphysics;
to obtain accurate predictions, it is critical
to
include weak magnetism corrections \citep{horowitz_02},
which increase the spread in mean energy by $\mathord{\sim}0.5\, \mathrm{MeV}$ \citep{fischer_18}
as they decrease the opacity for  $\bar{\nu}_e$,
and (especially at later phases) the effects of
nucleon interaction potentials on the charged-current rates \citep{martinez_12,roberts_12c}.
The difference in mean energy remains
remarkably constant with time in the most sophisticated
simulation codes throughout the accretion phase \citep{mueller_14,mirizzi_16,bruenn_16,seadrow_18}
and only decreases during the Kelvin-Helmholtz cooling phase
over time-scales of seconds.

Because the dominant opacities for electron-flavor neutrinos,
namely absorption and scattering on nucleons, strongly
depend on neutrino energy $ E_\nu$ with the  cross sections
scaling roughly as 
$\sigma\sim \sigma_0 ( E_\nu/m_e c^2)^2$
(where $\sigma_0=1.76 \times 10^{-44} \, \mathrm{cm}^2$), high-energy
neutrinos decouple further outside and 
the emerging spectra are therefore ``pinched'' with a steeper high-energy tail compared to  Fermi-Dirac spectra 
\citep{janka_89a,janka_89b,keil_03}. The
monochromatic neutrino numer flux $f_\nu$ for pinched spectra
can be conveniently parameterized by
a generalized Maxwell-Boltzmann spectrum in terms
of the mean energy $\langle E_\nu\rangle$ and a shape parameter $\alpha$
\citep{keil_03,tamborra_12},
\begin{equation}
\label{eq:pinch}
f_\nu\propto  E_\nu^\alpha
e^{-(\alpha+1)  E_\nu /\langle E_\nu\rangle},
\end{equation}
which has no particular motivation other than
the virtue of mathematical simplicity. Higher energy moments
$\langle E_\nu^n\rangle$,
\begin{equation}
\langle E_\nu^n\rangle=
\frac{\int f_\nu E_\nu^{n}\,\ud E_\nu}{\int  f_\nu E\,\ud E_\nu},
\end{equation}
of the distribution function given by
Equation~\ref{eq:pinch} can be calculated
recursively
in terms of the shape parameter $\alpha$ as:
\begin{equation}
\frac{\langle E_\nu^k\rangle}
{\langle E_\nu^{k-1}\rangle}
=\frac{k+\alpha}{1+\alpha}\langle E_\nu\rangle.
\end{equation}
\textbf{Table~\ref{tab:alpha}} lists typical
values for different stages based on the first and second energy moments 
from high-resolution spectra \citep{tamborra_12}.
Higher values of $\alpha$ indicate stronger pinching.

 \begin{table}[h]
 \tabcolsep7.5pt
 \caption{$\alpha$-parameters for high-resolution
 neutrino spectra from a $15 M_\odot$ progenitor}
 \label{tab:alpha}
 \begin{center}
 \begin{tabular}{@{}l|c|c|c@{}}
 \hline
 Species & Accretion phase & Early cooling phase & Intermediate cooling phase \\
        & ($261 \, \mathrm{ms}$) &  ($1016 \, \mathrm{ms}$) &
        ($1991 \, \mathrm{ms}$) \\
 \hline
 $\nu_e$ & 2.65$^\mathrm{a}$ & 2.90 & 2.92 
 \\
 \hline
 $\bar{\nu}_e$ & 3.13 & 2.78 & 2.61 
 \\
  \hline
 $\nu_x$ & 2.42 &  2.39 & 2.34 \\
 \hline
 \end{tabular}
 \end{center}
 \begin{tabnote}
 $^{\rm a}$ All values taken from the high-resolution
 case of \citex{tamborra_12}, Table~I.
 \end{tabnote}
 \end{table}

Spectrum formation is more complicated for 
$\nu_x$:
The emission and absorption of $\nu_x$
freeze out at higher densities and temperatures
than for $\bar{\nu}_e$, but
outside the ``number sphere'' where the
number flux of 
$\nu_\mu$, $\bar{\nu}_\mu$,
$\nu_\tau$, and $\bar{\nu}_\tau$ is set, $\nu_x$ can still exchange
energy with the medium via recoil in scattering reactions
on nucleons (which is the dominant energy transfer
mechanism during the accretion phase), electrons, and positrons
out to an ``energy sphere''. Since the average energy exchanged during
neutrino-nucleon scattering is small, the energy sphere lies somewhat inside
the surface of last scattering. The energy transfer can be sizable and reduce
heavy flavor neutrino luminosities by
$\mathord{\lesssim} 7\%$ in this scattering layer \citep{mueller_12a}.
As a result, the expected hierarchy
$\langle E_{\nu_e}\rangle<\langle E_{\bar{\nu}_e}\rangle<\langle E_{\nu_x}\rangle$
eventually changes to
$\langle E_{\nu_e}\rangle<\langle E_{\nu_x}\rangle<\langle E_{\bar{\nu}_e}\rangle$. The cross-over occurs earlier
for higher accretion rates; in massive progenitors
it may occur as early as
$\mathord{\sim}200 \, \mathrm{ms}$ after bounce.
The heavy-flavor neutrino spectrum remains less pinched than
that of $\bar{\nu}_e$
with $\alpha$-parameters of $\alpha \approx 2.4$ (see \textbf{Table~\ref{tab:alpha}}), however, so that one always has $\langle E_{\bar{\nu}_e}^2\rangle<\langle E_{\nu_x}^2\rangle$. 

Interestingly, simulations and
analytic considerations on the PNS surface
structure suggest that 
$\langle E_{\bar{\nu}_e}\rangle$ is roughly proportional
to the neutron star mass during the accretion phase.
For the EoS of \citex{lattimer_91}
with a bulk incompressibility modulus
of $220\, \mathrm{MeV}$, 
one finds \citep{mueller_14}
\begin{equation}
\label{eq:e_vs_m}
 \langle E_{\bar{\nu}_e}\rangle
\approx
10 \,\mathrm{MeV} (M/M_\odot).
\end{equation}
However, the proportionality constant is not independent of the nuclear EoS,
which can easily shift the mean energies
by up to several MeV during the later accretion phase. Even for a given high-density EoS, there is a scatter 
of $15\texttt{-} 20\%$ around the correlation $\langle E_{\bar{\nu}_e}\rangle
\propto M$ for different progenitors and epochs.

It has sometimes been  suggested that the electron flavor neutrinos also
provide a diagnostic for the onset of the explosion via a sudden drop of the luminosity around shock revival
because of the dependence on the mass accretion rate $\dot{M}$ \citep{wurm_12}.
This, however, is only an artifact of 1D explosion models
in which the explosion is triggered by hand.
Such a sudden drop is only associated with the infall of shell interfaces
in the progenitor (see Section~\ref{sec:progenitor}).
Different from 1D explosion models, multi-D models predict
a slow decline of $\dot{M}$ after shock revival \citep{mueller_14,bruenn_16,seadrow_18}
because there is an extended phase of concurrent mass ejection
and mass accretion. Although 3D models show
a faster decline of the accretion rate than 
2D models \citep{mueller_15b,mueller_17}, the  decline
is still drawn out over hundreds of milliseconds
(\textbf{Figure~\ref{fig:neutrino_signal}}, middle column).
It is essentially impossible to distinguish
whether such a gradual decline is due to shock revival or due
to the progenitor structure.

\subsection{Constraining Progenitor Properties}
\label{sec:progenitor}
In principle, flavor-dependent neutrino luminosities
and mean energies could be used to constrain the
time-dependence of the PNS mass $M$,
radius $R$, and the mass accretion rate $\dot{M}$
using Equations~\ref{eq:lum_x}, \ref{eq:lum_e}, and \ref{eq:e_vs_m}.
However, the amount of information that can actually be extracted
in this way from a future Galactic event will strongly depend on the
distance of the supernova from Earth, and both anisotropies in the
neutrino emission (Section~\ref{sec:multi_d}) and neutrino flavour
conversion introduce uncertainties in the interpretation
of the observed neutrino fluxes and spectra that cannot be easily
factored out.

A prominent feature in the time-dependent neutrino flux that likely survives even
with moderately high count rates is the drop in the
electron neutrino flavor luminosity that is associated with
the drop in $\dot{M}$ after the accretion of
the Si/O shell interface in many progenitor models (\textbf{Figure~\ref{fig:neutrino_signal}}, left and
middle column).
This drop is the consequence of a pronounced jump in entropy and density
at an active shell source with vigorous O burning at the onset of collapse.
 The infall time for the Si/O interface varies from
$\mathord{\sim}100\,\mathrm{ms}$
in low-mass progenitors to several hundreds of $\mathrm{ms}$
in high-mass progenitors.
Merely by timing the infall of this shell interface, one can therefore place important
constraints on the progenitor structure since the infall time 
$t_\mathrm{infall}$ of a shell
is directly related to its mass coordinate $m_\mathrm{if}$
and pre-collapse radius $r_\mathrm{if}$.
Approximately, one finds \citep{woosley_15b,mueller_16a}
\begin{equation}
t_\mathrm{infall}\approx
\sqrt{\frac{\pi^2 r_\mathrm{if}^3}{3 G m_\mathrm{if}}}
\end{equation}
although numerical simulations should be used to match the measured arrival time of the shell interface in practice.

\begin{figure}[htb]
\includegraphics[width=\linewidth]{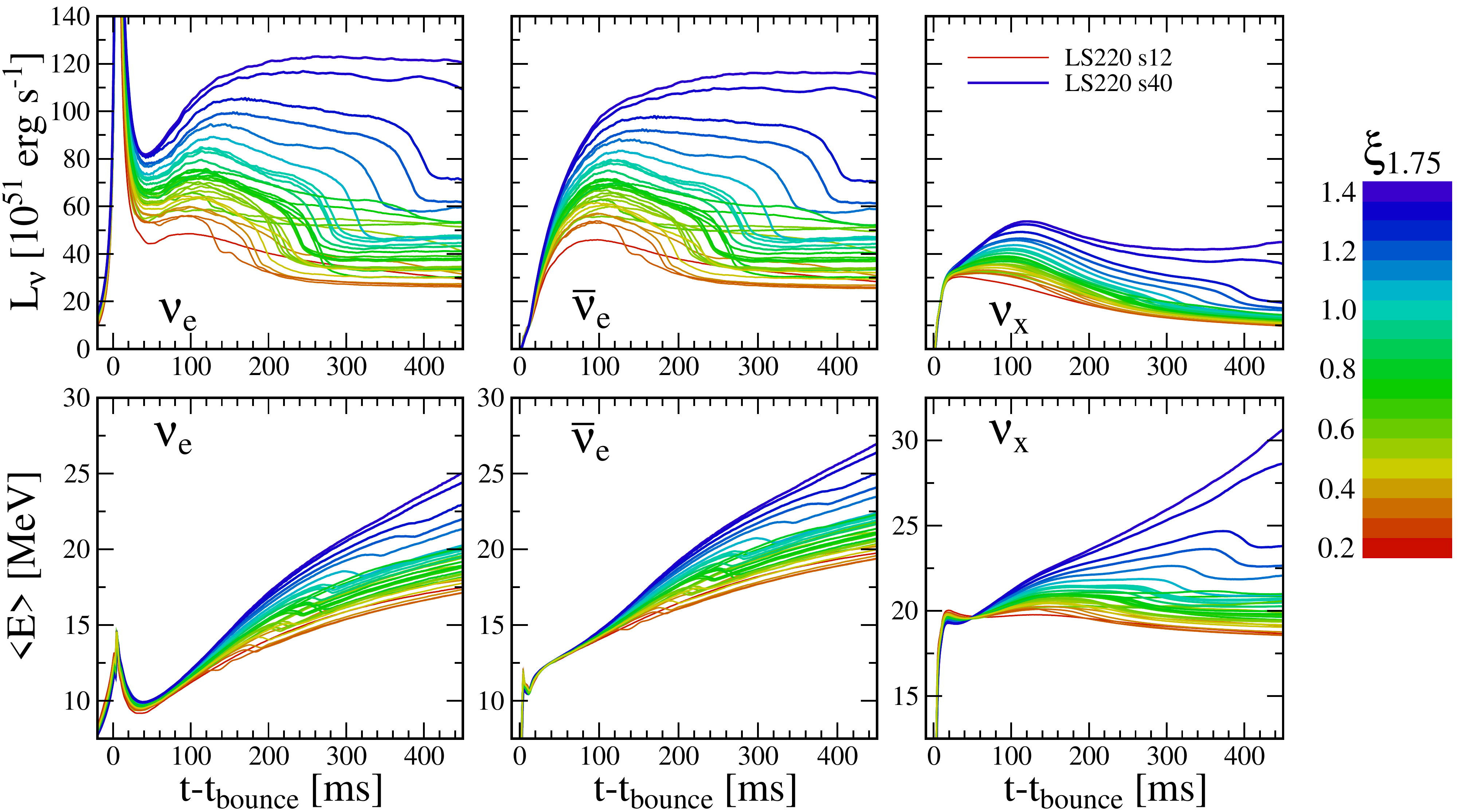}
\caption{Neutrino luminosities
$L_\nu$ and mean energies $\langle E \rangle$
for 32 progenitors with masses
between $12 M_\odot$
and $120 M_\odot$ obtained with the
LS220 EoS \citep{lattimer_91}. Colors
indicate the he progenitor
compactness $\xi_{1.75}$;
the trend towards higher luminosities and mean
energies for increasing $\xi_{1.75}$ is evident.
Figure from \citep{oconnor_13},
 \copyright The American Astronomical Society.
 Reproduced with permission.}
\label{fig:progenitors}
\end{figure}

Even without recourse to the detailed
time-dependence of the neutrino signal, one can still obtain  constraints on the progenitor
core structure from integrated count rates. This was already pointed
out after SN~1987A by
\citex{bruenn_87,burrows_88a} and recently reinvestigated by
\citex{oconnor_13,horiuchi_17} using large sets of progenitor models.
The study of \citex{oconnor_13} investigated the first
$\mathord{0.5} \, \mathrm{s}$ of post-bounce neutrino emission
of progenitors between $12M_\odot$ and $120 M_\odot$
in spherical symmetry 
(see \textbf{Figure~\ref{fig:progenitors}}). They showed that the energy
emitted in $\bar{\nu}_e$ varies by about a factor of
four across progenitors and is strongly correlated
with the
compactness $\xi_m$ of the progenitor, which is
essentially a normalized measure for the radius $r$ of a specified
mass shell $m$ \citep{oconnor_10},
\begin{equation}
\xi_m=\frac{m/M_\odot}{r/1000\, \mathrm{km}}.
\end{equation}
The progenitor variations in heavy-flavour neutrino emission
are less pronounced but still sizable. 
The compactness $\xi_{1.75}$ is a very good
predictor for the total pre-explosion neutrino emission \citep{oconnor_13};
 even in the ``worst case'' of a full swap
between $\bar{\nu}_e$ and $\bar{\nu}_x$ and even
with a present-day detector (Super-Kamiokande), the
cumulative inverse $\beta$-decay event count from
$\bar{\nu}_e$ is potentially a powerful diagnostic for the progenitor compactness. The
study of \citep{oconnor_13} also addresses degeneracies and uncertainties
that need to be overcome for a quantitative measurement of
the compactness, such as a possible drop of the accretion rate after shock revival, flavour conversion, rotation, and uncertainties
in the high-density EoS. Some of these degeneracies can be broken; specifically,  uncertainties in the EoS can be
eliminated by measuring both
the time-integrated flux and the mean energy of the detected time-integrated
spectrum from the pre-explosion phase \citep{oconnor_13}.

\begin{figure}[htb]
\includegraphics[width=0.92\linewidth]{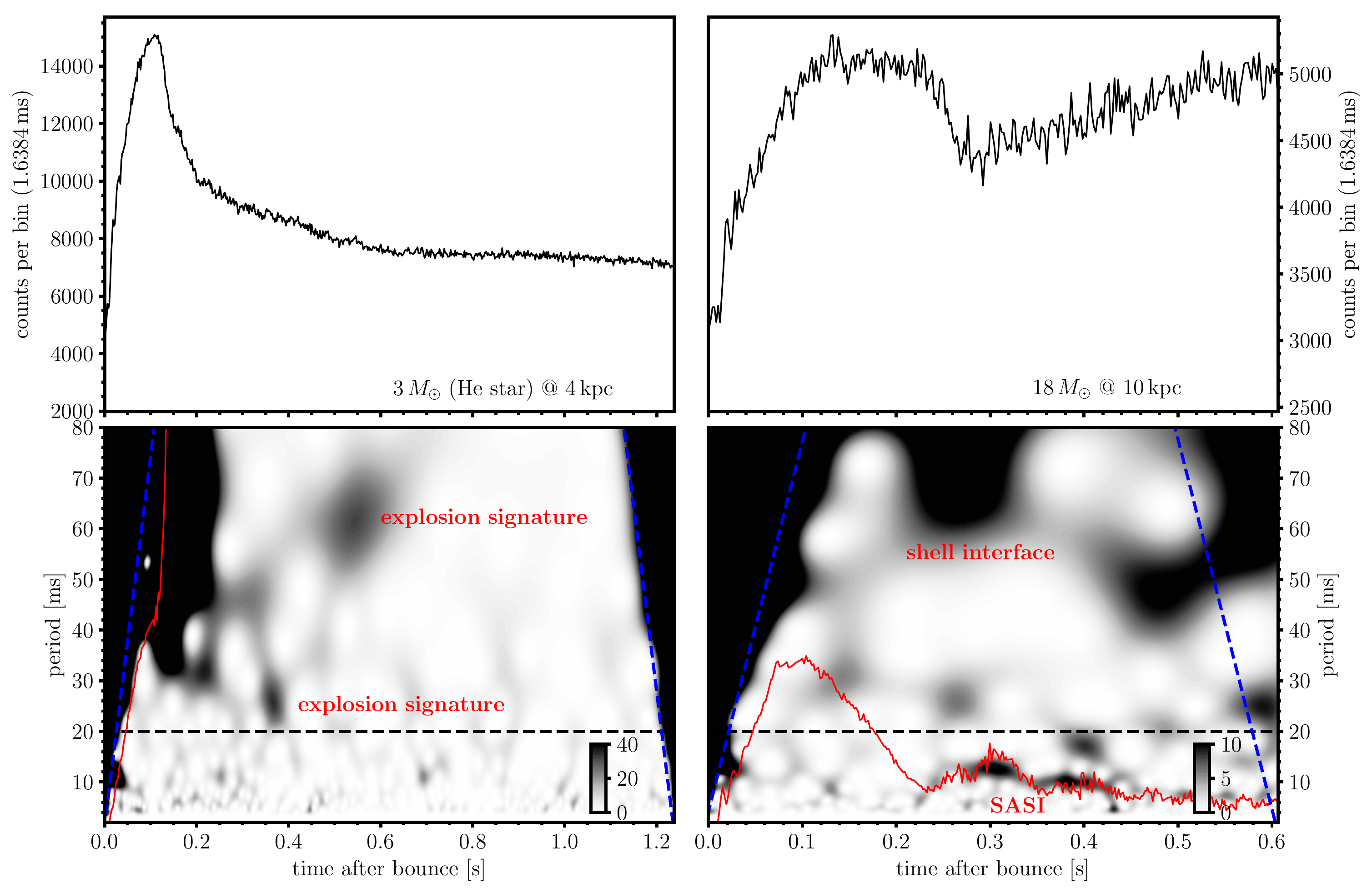}\\
\includegraphics[width=0.92\linewidth]{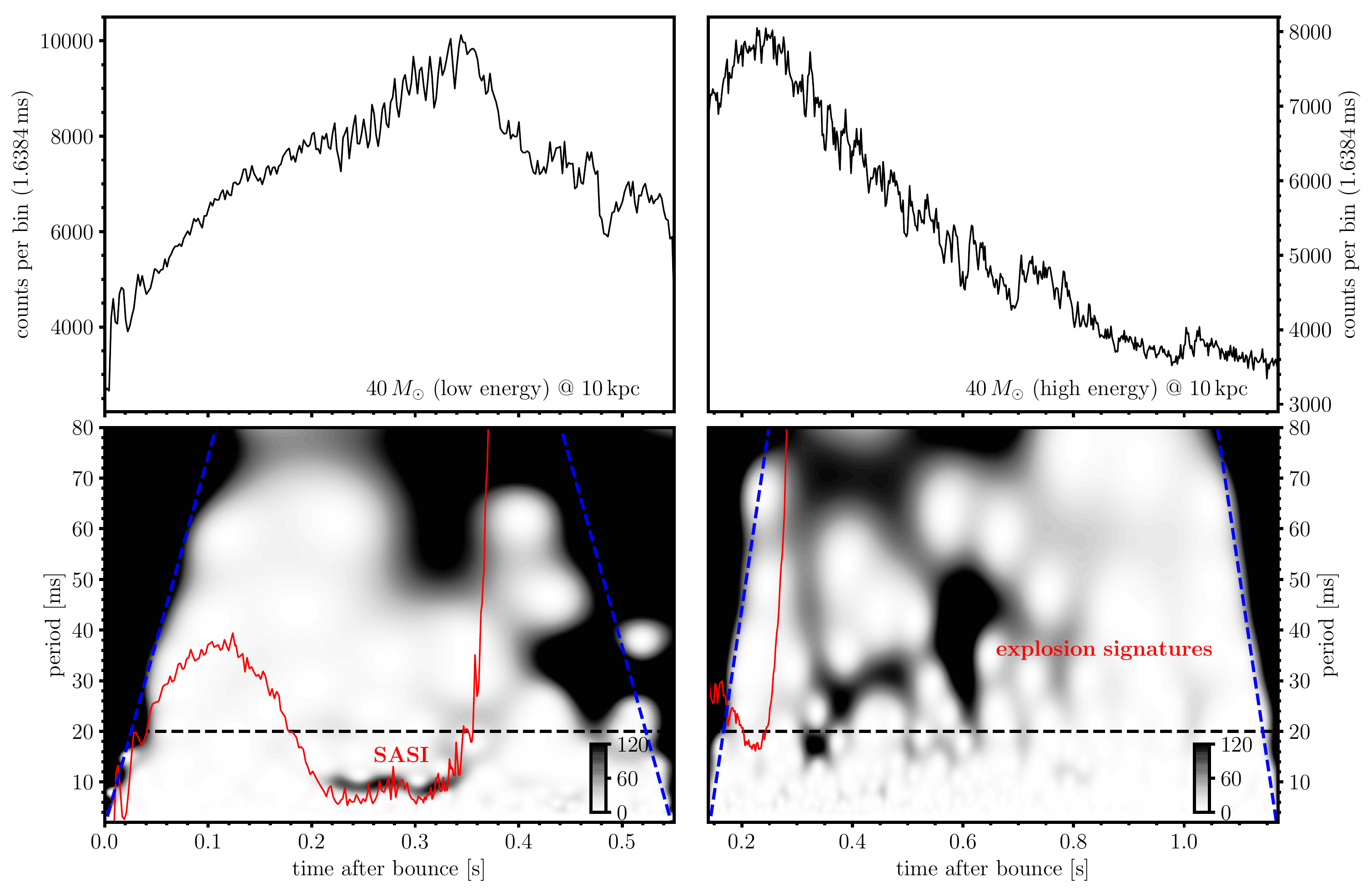}
\caption{Simulated IceCube count rates (including noise)
and their wavelet spectrograms for a $3 M_\odot$ He star
\citep{mueller_18b}
at $4 \, \mathrm{kpc}$,
the non-exploding $18 M_\odot$ model from
\citep{mueller_17}, the low-energy $40 M_\odot$ model
from \citex{chan_18}, and a high-energy $40 M_\odot$ case
in which an explosion was triggered early by
artificial pre-shock density perturbations
(all at $10 \, \mathrm{kpc}$). 
 The color in
the spectrograms indicates the signal-to-noise ratio;
dashed blue lines roughly demarcate the range of edge effects.
The models illustrate the
characteristic SASI fingerprint with its time-dependent
frequency, which is well fitted by Equation~\ref{eq:period} 
(red curves). Blobs at $\mathord{\gtrsim}20 \, \mathrm{ms}$ are
a smoking gun for the development of an explosion, but
need to be carefully distinguished from edge effects
and stripes from drops in the luminosity.}
\label{fig:modulation}
\end{figure}

\subsection{Imprint of Multi-Dimensional Fluid Flow on the Neutrino Signal}
\label{sec:multi_d}
One further obstacle for the interpretation of
observed neutrino fluxes and spectra is that the
neutrino emission will not be isotropic in the accretion
phase and early explosion phase. The emission of
electron flavour neutrinos is enhanced over accretion
hotspots \citep{marek_09}, and strong rotation
can lead to hotter spectra and enhanced luminosities
at high latitudes with somewhat different effects
on electron flavor and heavy-flavor neutrinos
\citep{ott_08_a,marek_09}. Moreover, recent simulations have
observed a strong global asymmetry in
the lepton number flux,  i.e., the difference between
the \emph{number} fluxes
of $\nu_e$ and $\bar{\nu}_e$, 
which is
connected to a slowly evolving, low-mode instability
in the PNS convection zone
\citep{tamborra_14a,janka_16,glas_18b,oconnor_18,powell_18}. 
After independent corroboration with many
neutrino transport codes, there is little
doubt that this
``Lepton-number Emission Self-sustained Asymmetry''
(LESA, \citealp{tamborra_14a}) is not
a numerical artifact, but the phenomenon is still not fully
understood. There are some indications that it may
be nothing more but a manifestation of buoyancy-driven
PNS convection whose peculiarities -- in particular the slowly evolving dipole mode in
the lepton number distribution -- are related to the presence of partially stabilizing lepton number gradients and diffusive transport
\citep{janka_16,glas_18b,powell_18}, but a rigorous
theory of the LESA is still lacking. 

Uncertainties from orientation effects
are difficult to control for, but thankfully, they
may be of a modest scale (except for the orientation
effect on the lepton number flux) and dwarfed by uncertainties related to flavor conversion. In non-rotating
3D models, variations in the neutrino fluxes
due to wandering accretion downflows remain
below $\mathord{\sim} 10\%$ and tend to average
out over time. Systematic errors
from latitudinal variations of the neutrino emission
from rotating PNSs
are more difficult to eliminate, but 
constraints
on the progenitor rotation from the gravitational
wave signal may help \citep{abdikamalov_14}.

Intriguingly, the modulation of the neutrino emission by asymmetric accretion onto the PNS can even be used to probe the dynamics of multi-D flow in the supernova core. Already the 
simulations of \citex{ott_08_a,marek_08} revealed
that the prominent sloshing motions of the SASI
in 2D are clearly mirrored in the neutrino signal
for appropriate observer directions. Subsequent
studies established that these modulations
are detectable by IceCube \citep{lund_10,lund_12,brandt_11,mueller_14,walk_18},
HyperK \citep{tamborra_13,tamborra_14b,walk_18},
DUNE, and JUNO \citep{seadrow_18}. For SASI-dominated
models, these accretion-induced modulations in the 
neutrino signal are every bit as strong
in 3D as in 2D \citep{tamborra_13,tamborra_14b,walk_18} and visible to $10\texttt{-} 20 \,\mathrm{kpc}$.
Although they can be considerably less pronounced 
in models without SASI, they remain
detectable with instruments like IceCube
albeit only to a few kpc \citep{lund_12}.
Various studies demonstrated that the frequency
spectrum of the modulations shows clearly identifiable
peaks that can be used to infer something of
a typical, time-averaged SASI
frequency \citep{lund_10,tamborra_13,tamborra_14b,walk_18}. Such peaks would not only serve as a smoking
gun for SASI activity in the supernova core, but even
quantitatively constrain the key parameters that determine its frequency, namely the shock radius
and the radius of maximum deceleration
\citep{foglizzo_07}, which is very similar
to the PNS radius.

\begin{textbox}[h]\section{Challenges of neutrino transport in three dimensions}
Capturing the effects of anisotropic neutrino emission on the dynamics
and the observable neutrino signal requires multi-D neutrino
transport.  Retaining the full six-dimensional phase space dependence
of the radiation field using discrete ordinate
\citep{nagakura_16} or Monte Carlo methods \citep{abdikamalov_12}
is still impractical for dynamical simulations over long time scales in
3D. Thus, various approximations for multi-D transport are currently
used. The \emph{ray-by-ray                                                                                                       
approximation} \citep{buras_06a,mueller_10,bruenn_18} provides a
straightforward way to generalize sophisticated transport algorithms
for spherical symmetry to multi-D by retaining only a parametric
dependence of the radiation field on the angular coordinates in real
space. It overestimates anisotropies in the neutrino emission, which
can precipitate explosions in 2D
\citep{skinner_16,just_18}, but has little impact on the dynamics in 3D \citep{glas_18a}. Ray-by-ray
simulations can be post-processed to obtain more accurate fluxes and
spectra for any given observer direction.
\emph{Flux-limited diffusion} without the ray-by-ray approximation
artificially smears out anisotropies in the radiation field \citep{ott_08_a}.  \emph{Two-moment methods                          
using an analytic closure} have emerged
\citep{just_15,roberts_16,kuroda_16,skinner_18,oconnor_18a} as a popular approximation that
captures the anisotropies in the radiation field quite
well.  As far as the observable signatures of anisotropic neutrino
emission are concerned, there is no fundamental disagreement between
post-processed ray-by-ray models, Boltzmann transport, and two-moment
transport.
\end{textbox}

Given sufficiently violent fluid motions in the 
supernova core, the modulation of the neutrino signal 
can even be strong enough to deduce more detailed, time-dependent information on the dynamics
from spectrograms of the signal \citep{mueller_14,walk_18}. This has been worked
out most fully in \citex{mueller_14} for
a range of exploding and non-exploding 2D models with the help of wavelet spectrograms of simulated
signals in IceCube. 
\textbf{Figure~\ref{fig:modulation}} illustrates the diagnostic
power of signal spectrograms using more recent
3D models, which were analyzed using the same
assumptions as in \citex{mueller_14}, i.e.,
only MSW flavor conversion in the normal mass hierarchy and a simple detector model for IceCube.
For SASI-dominated models of $18 M_\odot$  
\citep{mueller_17}
and $40M_\odot$  \citep{chan_18}, the spectrogram shows significant  power at periods of $\mathord{\sim} 10\, \mathrm{ms}$ at late post-bounce times.
The time-dependent period
$T_\mathrm{SASI}$ of the SASI peak is
well described by
\begin{equation}
\label{eq:period}
    T_\mathrm{SASI}
    =19 \, \mathrm{ms}
    \left(\frac{r_\mathrm{sh,min}}{100 \, \mathrm{km}}\right)^{3/2}
    \ln \left(\frac{r_\mathrm{sh,min}}{R}\right),
\end{equation}
in terms of the minimum shock radius
$r_\mathrm{sh,min}$ and the  PNS radius $R$ \citep{mueller_14}. Combined
with information on the PNS
mass and radius from the neutrino luminosities and mean
energies, and, under favorable circumstances,
gravitational waves \citep{mueller_13},
Equation~\ref{eq:period} can in principle be used
to constrain the shock trajectory.

The accretion-induced signal modulations also provide
a telltale sign for the onset of the explosion,
namely a
shift of power beyond
periods of $\mathord{\sim}20 \, \mathrm{ms}$
\citep{mueller_14}. Furthermore, 
\citex{mueller_14} found  small bursts in the emission of $\nu_e$
and $\bar{\nu}_e$ due to episodic fallback in some
models; similar phenomena can be
seen in the 2D models of \citex{seadrow_18}. However, these signatures of the explosion
are somewhat exaggerated by symmetry artifacts in 2D models, where the accretion downflows hit the PNS with higher velocities  and the accretion rate fluctuates
considerably more than in 3D after shock revival
\citep{mueller_15b}. In 3D, the accretion-induced
modulation of the neutrino signal tends to be much
milder even when there is ongoing accretion
after shock revival. In the case of the $3M_\odot$ helium
star model of
\citex{mueller_18b}, the simulated
spectrogram of the IceCube signal only shows 
the characteristic wavelet power at long periods for
a supernova distance of
$4 \, \mathrm{kpc}$ (top left panel in \textbf{Figure~\ref{fig:modulation}}). Strong explosion signatures survive
only in massive progenitors with high accretion
rates after shock revival, for example
in another, more energetic $40 M_\odot$ model similar to
that of \citex{chan_18}.

Recently, \citex{walk_18} considered the  modulations
of the neutrino emission as a probe of progenitor rotation.
They found that rotation changes the amplitudes and the direction-dependence of the signal modulation due to a number of effects
that depend on the rotation rate and whose interplay
appears to be quite intricate. They identify distinct features
in the modulation spectra and spectrograms of rotating models,
such as secondary peaks above the SASI frequency. Our understanding
of these rotational effects is  somewhat sketchy at present, however,
and their diagnostic potential still needs to be investigated further.

\subsection{Black Hole Formation and Phase Transition Signatures}
\label{sec:bh_formation}
The time-dependent neutrino fluxes could provide 
further clues about the dynamics in the supernova 
core in the case of black hole formation,
which would lead to
a sharp cut-off of neutrino
fluxes during the first seconds after bounce. Such a cut-off would
likely be preceded just by a gradual rise of the neutrino luminosities
and mean energies. Although
some calculations \citep{fischer_09} indicated
that black hole formation could be associated with a noticeable rise
in the heavy-flavour luminosities and 
several MeV in the mean energies 
of  $\nu_\mu$ and $\nu_\tau$
as their neutrinosphere contracts strongly when the PNS approaches the critical mass for collapse, this strong rise
disappears when the energy exchange with the medium in the scattering
layer (see Section~\ref{sec:bulk_parameters}) is taken into account
\citep{huedepohl_phd,mirizzi_16}. 

The detection of such a cut-off could help resolve a number of questions
in nuclear physics and astrophysics. In principle,
the neutrino emission and the time of black hole formation is sensitive to the EoS \citep{sumiyoshi_07}; but they also depend on
the progenitor \citep{fischer_09}, and it remains to be
seen how well these factors can be disentangled by
combining neutrino and electromagnetic observations.
  The astrophysical implications of a timed observation
of black-hole formation might be even broader. 
Simulations \citep{chan_18} as well as observational evidence
from the composition of metal-poor stars \citep{keller_14}
and  of companions in HMXBs
\begin{marginnote}[]
\entry{HMXB}{High-mass X-ray binary}
\end{marginnote}
\citep{podsiadlowski_02}, and from the kinematics of some HMXBs
\citep{podsiadlowski_02,repetto_12} suggest that black hole formation can sometimes occur \emph{after} shock revival due to fallback. A cut-off in the neutrino flux together with an explosion
of sufficient energy would be a direct proof for this scenario.

The neutrino signal may also reveal phase transitions at high densities in the PNS.
More conservative scenarios
of a late phase transition in the
cooling will be discussed in Section~\ref{sec:pns_cooling},
but models for an early first-order phase transition at relatively low density
have also been proposed: 
In this scenario, the phase transition
leads to a second collapse of the PNS and the formation of a secondary shock that could trigger an explosion \citep{sagert_09,fischer_18}. The formation of such a secondary shock
would lead to a small, secondary neutrino burst. Different from the
neutronization burst, this burst would be seen in all flavors
and for both neutrinos and antineutrinos.
$\bar{\nu}_e$ would be most abundantly emitted,
however, because the
hot $\beta$-equilibrium
 is suddenly shifted to higher $Y_e$ by shock heating, so that
the shocked matter protonizes. The signal of such
a second burst in IceCube and Super-Kamiokande
was analyzed by
\citex{dasgupta_10}: For the  EoS
of \citex{sagert_09}, the brief increase of the detector
count rates by a factor of several would serve as a clear
fingerprint of the phase transition for a Galactic
supernova even at a distance
of $20 \, \mathrm{kpc}$. However, it remains to be seen 
whether such a secondary burst can be distinguished from temporal modulations
of the neutrino signal by wandering accretion downflows and fallback, 
if the phase transition leads only to a weaker second bounce than
in the models of \citex{sagert_09,fischer_18}. 
Moreover, the viability of phase transition models is already
limited by a number of other constraints. For example, the
EoS originally used by \citex{sagert_09} is incompatible
with the highest measured neutron stars masses
\citep{antoniadis_16}, and the light curves
from powerful explosions driven by a phase-transition
fall in the category of superluminous supernovae
or peculiar SN1987A-like type~IIPs
\citex{fischer_18}, which places significant
limits on the prevalence of this explosion scenario.

\section{THE KELVIN-HELMHOLTZ COOLING PHASE}
\label{sec:pns_cooling}
As accretion ceases, diffusive
transport from within the PNS becomes the only source
of neutrino emission in the Kelvin-Helmholtz cooling phase. As a result, the electron flavor
and heavy flavor luminosities become relatively similar
(\textbf{Figure~\ref{fig:neutrino_signal}}, right column).
The luminosities decrease roughly exponentially with a decay
time-scale of seconds. Despite the energy
loss, the surface temperature and hence the mean energies
of the neutrinos still increase 
for $\mathord{\sim} 1\,\mathrm{s}$
\citep{huedepohl_10,mirizzi_16} due to the contraction of the PNS. 

The spectra of the different neutrino species remain different
with $\bar{\nu}_e$ maintaining higher mean energies than
$\nu_e$, such as to maintain a net lepton number flux out of the
PNS. Modern simulations
of the Kelvin-Helmholtz cooling phase show that the 
mean energies of $\nu_x$ remain below
those of $\bar{\nu}_e$ due to recoil energy transfer 
in the scattering layer \citep{huedepohl_10}. As the
PNS cools, the absolute differences in mean energy
between the neutrino species shrink.

The demarcation between the accretion phase and the
cooling phase is not a sharp one; there is a rather a
gradual transition in the character of the neutrino emission.
Although recent 3D explosion models show ongoing accretion
at some level over time scales of seconds
\citep{mueller_17,mueller_18b}, the neutrino emission
in these 3D models already exhibits some  features of the cooling phase 
a few hundred milliseconds after shock revival, i.e.,
similar luminosities of all flavors, and little
short-term variation due to variable accretion downflows.

\subsection{Sensitivities of the Neutrino Emission}
One of the key parameters determining the neutrino emission during
the Kelvin-Helmholtz cooling phase is the neutron star binding energy
$E_\mathrm{bind}$, which depends on the PNS mass
and the EoS
(see Equation~\ref{eq:ebind}).
The major fraction of $E_\mathrm{bind}$ 
is radiated away after shock revival when the luminosities of all neutrino
flavors have become similar. Hence the total energy emitted
in (anti-)neutrinos of all three flavors are also similar (equipartition), and  serve
as a measure for $E_\mathrm{bind}$ that
is not too strongly affected by uncertainties in flavor conversion.
For example, \citex{huedepohl_10} find good
equipartition in their cooling models for an electron-capture supernova progenitor. For more massive
progenitors with extended
accretion, equipartition does not hold quite as well. As shown  in
\textbf{Table~\ref{tab:s18_equipartition}},
extrapolating the neutrino emission in the 
$18 M_\odot$ explosion model of \citex{mueller_17} puts the total energy in each neutrino species
to within $20\%$ of the equipartition value $E_\mathrm{bind}/6$.

 \begin{table}[ht]
 \tabcolsep7.5pt
 \caption{Energy budget for the different neutrino species
 in an $18 M_\odot$ star}
 \label{tab:s18_equipartition}
 \begin{center}
 \begin{tabular}{@{}l|c|c|c|c@{}}
 \hline
 Species & Energy up to $t=2.4 \, \mathrm{s}$ 
 ${}^\mathrm{a}$ & Residual energy${}^\mathrm{a}$  & Total & Relative to\\
        & ($10^{52} \, \mathrm{erg}$) &  ($10^{52} \, \mathrm{erg}$) &
        ($10^{52} \, \mathrm{erg}$) & equipartition\\
 \hline
 $\nu_e$ & 5.2  & 3.3 & 8.5  & +20\%\\
 \hline
 $\bar{\nu}_e$ & 4.8 & 3.3 & 8.1 & +16\% \\ 
  \hline
 $\nu_x$ & 3.1 & 3.3 & 6.4 & -9\%\\
 \hline
Total & 22.4 & 19.6 & 42 & \\
\hline
 \end{tabular}
 \end{center}
 \begin{tabnote}
 $^{\rm a}$ Obtained by numerical integration of the luminosities
 from  \citex{mueller_17}.
\\ 
 $^{\rm a}$ Obtained assuming equipartition after
 $2.4 \, \mathrm{s}$ after bounce and a binding
 energy of $4.2\times 10^{53}\, \mathrm{erg}$
 for a putative neutron star mass of $1.67 M_\odot$.
 \end{tabnote}
 \end{table}

The time-dependence of the neutrino luminosities and mean energies
is sensitive to various factors. Different from the accretion phase, the stratification, thermodnynamic conditions, and transport coefficients deep in the PNS 
now play a key role in shaping the neutrino emission as the slow evolution of the
interior by neutrino diffusion makes itself felt in the neutrinospheric
conditions.  This makes the cooling phase a better laboratory for uncertain
nuclear physics well above saturation density. However, it is not trivial
to extricate the underlying physics from the neutrino fluxes and spectra since different effects and nuclear physics
parameters can affect the neutrino emission in a similar way.

One of the most important factors regulating the duration of the cooling
phase are the neutrino opacities around and above saturation density. 
In the relevant equilibrium diffusion regime, it is the \emph{total} Rosseland-averaged
opacity that determines the energy and lepton number flux, and hence the most critical opacities
are those for charged-current absorption and neutral-current scattering.
At high densities these are strongly affected by in-medium (correlation) effects
\citep{burrows_98,burrows_99,reddy_99,horowitz_17}. 

That in-medium effects can significantly change the
PNS cooling time scale was already realized decades ago \citep[e.g.][]{sawyer_95,keil_95}. The first modern cooling models 
\citep{huedepohl_10,huedepohl_phd,mirizzi_16} including nucleon correlations following
the RPA
\begin{marginnote}[]
\entry{RPA}{Random Phase Approimation}
\end{marginnote}
 framework of \citex{burrows_98,burrows_99} predicted considerably
shorter cooling times of the order of seconds due
to the reduced opacities at high densities, rather than tens of seconds
in older models \citep{keil_95,pons_99,fischer_10}.
However, whereas measurements can be used to constrain correlation effects
at moderate densities using the virial approach \citep{horowitz_17}, considerable
uncertainties remain in the relevant high-density opacities well above saturation density.
Moreover, many-body effects do not invariably lead to a reduction of the
opacities, and this can
also have a noticeable effect on the neutrino emission from the cooling phase
even though enhanced opacities may only apply in a thin layer near the
PNS surface. For example, \citex{horowitz_16}
consider the effect of an enhanced neutral-current scattering opacity
due to the formation of nuclear pasta and find a delay of the cooling with significantly
increased neutrino luminosities and mean energies at late times.

The high-density EoS also affects the emission
from the cooling phase in other ways. Differences in
neutron star radius and the location of $\beta$-equilibrium
translate into differences in the gradients of
the temperatue and neutrino chemical potential that drive
the diffusive energy and lepton number flux, and also
affect the Rosseland-averaged neutrino opacities
as these depend on density, temperature, and lepton
number. In addition, the EoS
affects the extent of the convective region
inside the PNS during the cooling
phase \citep{roberts_12b,mirizzi_16}.
The resulting EoS-dependence of the 
cooling and deleptonization time scale is
non-trivial, and  while there appears to  be a weak trend towards
shorter cooling time scales for stiffer EoSs
\citep{huedepohl_phd,mirizzi_16},
no hard-and-fast rule can be given.
Moreover, the cooling time scale also depends on the
PNS mass $M$ with a a trend towards a longer
cooling time for higher $M$ \citep{huedepohl_phd}.

\begin{figure}[htb]
\includegraphics[width=0.6\linewidth]{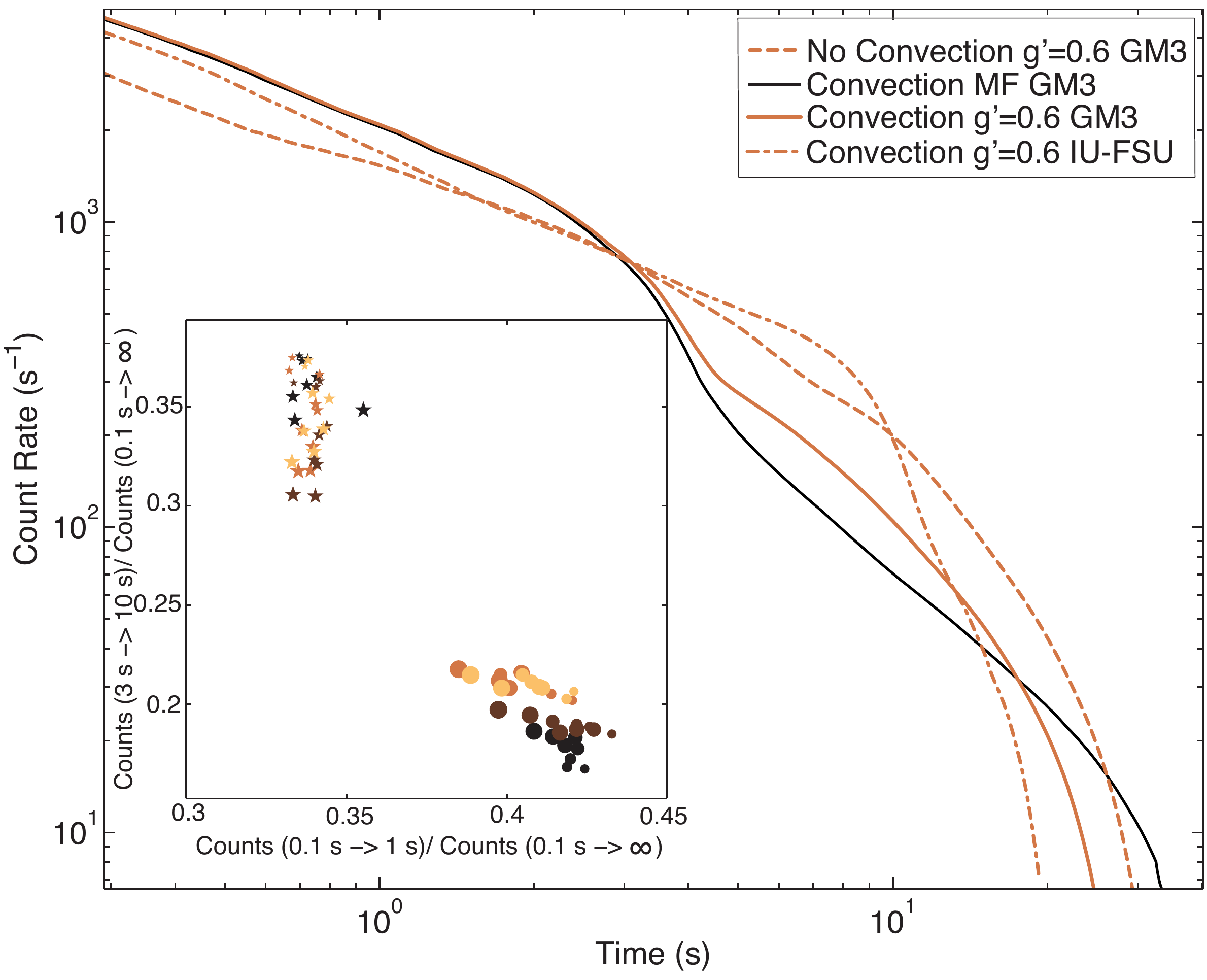}
\caption{Count rates in SuperKamiokande for PNS
cooling models with $M=1.6M_\odot$ with and without mixing-length convection
for different EoSs (GM3 vs.\ IU-FSU), and different
values of the Migdal parameter ($g'=0.3$ vs.\ $g'=0.6$) in the calculation of RPA opacities. The break
of the luminosities in the convective models indicates
the disappearance of the PNS convection zone, which occurs at different times for the two EoSs.
The inset shows that the fraction of counts
after $3\, \mathrm{s}$ and before $1\,\mathrm{s}$
can be used to separate the two EoSs
(circles: GM3, stars: IU-FSU) for a number
of cooling models with PNS masses between
$1.2 M_\odot$ and $2.1 M_\odot$. Figure
from \citep{roberts_12b},
\copyright APS. Reproduced with permission.}
\label{fig:pns_conv}
\end{figure}

 One signal feature whose connection
to the underlying EoS physics has been explained in some detail
is a break in the neutrino luminosity that occurs in models in
which the convection zone disappears during the cooling phase \citep{roberts_12b} as shown in \textbf{Figure~\ref{fig:pns_conv}}.
The disappearance of the convection zone, and hence the break in the neutrino light curves can be related
to the derivative 
$\pd E_\mathrm{sym}/\pd \rho$ of the
nuclear symmetry energy $E_\mathrm{sym}$;
large values of $\pd E_\mathrm{sym}/\pd \rho$ favor the earlier
termination of convection because they imply a stronger stabilizing
effect of the negative lepton number gradients under the conditions
encountered during the cooling phase (i.e.\ low temperatures and
$Y_e$ close to $\beta$-equilibrium).

There is also the possibility that some
phase transition occurs during the PNS
cooling case. Different from the phase 
transition scenario in 
Section~\ref{sec:bh_formation}, such 
a phase transition would not be
triggered by additional accretion, but merely by the contraction of the PNS during the cooling phase and the ongoing deleptonization. Studies that
investigated the appearance of hyperons or a phase transition to a kaon condensate or quark matter during the cooling phase determined that 
the cleanest signature for such a
phenomenon would be a cut-off of
the neutrino flux in case the phase
transition triggers collapse
to a black hole at late times; otherwise the
effect on the neutrino light curves is small
\citep{pons_99,pons_01a,pons_01b}. However,
the scenario of delayed collapse needs
to be revisited with updated
EoS models in the light of
more recent constraints on the high-density
EoS, such as better limits on the maximum neutron star mass \citep{antoniadis_13},
 and on neutron star radii \citep[e.g.][]{bauswein_17}.

\subsection{Constraints on Exotic Energy Loss Channels}
\label{sec:exotic}
It has long been realized that the time-integrated neutrino flux
and the duration of the neutrino signal can be used to place
constraints on the emission of hypothetical particles
such as axions, sterile right-handed
neutrinos \citep{raffelt_96},
and Kaluza-Klein gravitons \citep{hannestad_01} 
that would carry away
a sizable fraction of the PNS binding energy. Especially cooling by axions has been
studied extensively: The detection of neutrinos from SN~1987A already
helped to place an upper limit
on the axion mass $m_a$;
initial estimates of an upper limit of $m_a\lesssim 10^{-3}\,\mathrm{eV}$
 \citep{turner_88,raffelt_88,mayle_88}
have since then been weakened 
to $m_a\lesssim 10^{-2}\,\mathrm{eV}$
(with the precise limit depending somewhat on the axion model)
because of many-body effects that modify the axion cooling rate
\citep{janka_96b,keil_97}. Prospects
for better bounds on the axion mass from a Galactic supernova
have recently been investigated
by \citex{fischer_16}, but their results do not promise
substantially better bounds ($m_a\lesssim 10^{-2}\,\mathrm{eV}$)
from the neutrino light curves. 

For some of these exotic particles,
there does not appear to be much room for
 improved bounds due to better detection statistics, since this would require
tracking down extra energy loss that only amounts to a small
fraction of the neutron star binding energy; and at this level
uncertainties in the neutron star mass, radius, and equation
of state can no longer be ignored.

In some scenarios, specifically those involving sterile neutrinos
\citep[e.g.][]{esmaili_14} or non-standard neutrino interactions
\citep[e.g.][]{esteban_07b}, one would expect clearer signatures in the neutrino emission
(not necessarily during the cooling phase) from a nearby supernova with sufficiently high neutrino count rates and sufficient
temporal resolution. For example, there is the possibility
of energy-dependent jumps in the observed neutrino fluxes as the conditions for
flavor conversion and hence the survival
probabilities of observable flavors change \citep{esmaili_14}.

\subsection{Shock Propagation Effects}
Beyond serving as a probe for the conditions in the PNS,
the signal from the cooling phase may also provide clues about the explosion dynamics.
For normal iron-core progenitors with shallow density profiles, the
shock traverses the MSW resonance regions during the cooling phase,
and as a result MSW flavor conversion becomes non-adiabatic
\citep{schirato_02,fogli_03,tomas_04}. Such a change in MSW flavor conversion
could lead to detectable changes in the neutrino spectra or other
convenient measures that are sensitive to flavor conversion like
the ratio of charged-current to neutral-current event rates
\citep{kneller_08}. Similarly, MSW flavor conversion will be affected
by the formation and propagation of a reverse shock in the wake
of shock deceleration \citep{tomas_04}. 
This prospect of such a late-time signature from shock propagation 
in the neutrinos signal is intriguing,
but there are also several complications.
Since multi-D fluid instabilities
play a major role already during the phase of shock revival and
also later on as the shocked shells become unstable to Rayleigh-Taylor
mixing, 
the phenomenology of MSW
flavor conversion is affected by stochastic fluctuations
of the density (and hence of the matter potential) behind the shock.
These can modulate and even suppress oscillation signatures \citep{friedland_06,fogli_06,kneller_10}. Our incomplete understanding
of collective flavor conversion also presents a problem.
Finally, since much of the work on the signatures
of shock propagation and turbulence still assumes larger
spectral differences between flavors than obtained in 
modern simulations of the cooling phase, many findings
on the associated neutrino signatures deserve to be revisited at some point.

\begin{summary}[SUMMARY POINTS]
\begin{enumerate}
\item 
Detailed time-dependent information on the neutrino fluxes and spectra as expected
from current and future neutrino detectors
with complementary designs
is the key to exploiting neutrinos as a diagnostic of core-collapse supernovae.
\item 
The $\nu_e$-burst and the early post-bounce phase can provide a handle on  
the supernova distance, the neutrino mass ordering, and possibly on the progenitor structure in the case of an electron-capture supernova. The rise of the signal also provides precise timing information, which is of relevance for gravitational wave detection.
\item 
During the accretion phase, flavor-dependent
fluxes and spectra would help place constraints on the time-dependent PNS surface temperature, mass, and radius and
the accretion rate via the Stefan-Boltzmann
law, the excess accretion luminosity of $\nu_e$ and $\bar{\nu}_e$, and the
relation $\langle E_{\bar{\nu}_e}\rangle\propto M$.
\item 
Barring uncertainties concerning
flavor transformation, the time-integrated neutrino emission
during the accretion phase can be 
used to infer the progenitor
compactness, and the position of
the Si/O shell interface can be constrained using the characteristic
drop in electron flavor luminosity.
\item 
The modulation of the neutrino emission
by the time-varying accretion flow onto the PNS can be used to infer the presence of the SASI and measure the time-dependence of the SASI frequency, which is related to the shock radius. Temporal modulations with periods $\mathord{\gtrsim} 20\,\mathrm{ms}$ serve as indicator for a developing explosion.
\item
The neutrino emission from the Kelvin-Helmholtz cooling phase serves as a probe for the structure and microphysics (high-density EoS, opacities) of the PNS interior. The time-integrated flux constrains the neutron star binding energy and is relatively robust against uncertainties from flavor conversion because of approximate flavor equipartition in the 
cooling phase.%
\end{enumerate}
\end{summary}


\begin{issues}[FUTURE ISSUES]
\begin{enumerate}
\item Although supernova simulations
have matured considerably, there will still be room in the coming decade for further technical improvements, a better exploration of
parameter space (ideally by means
of 3D simulations from collapse into the cooling phase), and broader replication of results by different groups in order to understand the phenomenology of supernova neutrino emission. 
\item Flavor conversion remains a thorny issue for inferring supernova physics from the neutrino signal. If it turns out that fast flavor conversion can occur in the neutrinospheric region already during the accretion phase, this would pose a serious challenge for supernova modelling and force us to revise much of the current neutrino signal predictions.
\item While we focused on the one-in-a-lifetime chance of a Galactic
supernova in this review, there is also
the possibility of exploiting
the diffuse supernova neutrino background (DNSB),
see \citep{beacom_10,mirizzi_16,lunardini_16}
for extensive reviews. The DNSB will provide
complementary information, in particular
on the fraction of failed supernovae
\citep{lunardini_09} and hence on
the mass range for succesful explosions \citep{horiuchi_18}.
\end{enumerate}
\end{issues}

\section*{DISCLOSURE STATEMENT}
The author is not aware of any affiliations, memberships, funding, or financial holdings that
might be perceived as affecting the objectivity of this review. 

\section*{ACKNOWLEDGMENTS}
I wish to thank
E.O'Connor 
and L.~Roberts for permission
to use Figures~\ref{fig:progenitors} and
\ref{fig:pns_conv}, respectively. This work was supported by the Australian Research Council through
ARC Future Fellowship FT160100035. This research was undertaken with the assistance of
resources obtained via NCMAS and ASTAC  from the National Computational Infrastructure (NCI), which
is supported by the Australian Government and was supported by
resources provided by the Pawsey Supercomputing Centre with funding
from the Australian Government and the Government of Western
Australia. 

%

\bibliography{main}

\end{document}